\documentclass[twoside]{dis04}

\usepackage{amsmath}
\def\runauthor{G.~P.~Salam}
\def\shorttitle{CAESAR: Computer Automated Resummations}

\usepackage{xspace}
\usepackage{hyperref}

\newcommand{\ie}{i.e.\ }
\newcommand{\eg}{e.g.\ }

\newcommand{\as}{\alpha_s}              

\newcommand{\NC}{N_c}

\newcommand{\cF}{{\cal F}}
\newcommand{\ee}{e^+e^-}
\newcommand{\caesar}{\textsc{caesar}\xspace}

\begin{document}

\title{CAESAR: Computer Automated Resummations}

\author{Gavin~P.~Salam}

\address{LPTHE, Universities of Paris VI \& VII and CNRS,
75252 Paris 75005, France}

\maketitle


\abstracts{This talk gives a brief discussion of the motivations and
  principles behind computer automated expert semi-analytical
  resummation (\caesar) for QCD final states.  }



Global properties of energy-momentum flow in final-states, such as
event shapes and jet-resolution thresholds, offer a good compromise
between simplicity and sensitivity to the dynamics of QCD. Thanks to
the former, it has been possible to make a wide range of theoretical
predictions for them, including both fixed-order and resummed
perturbative calculations as well as non-perturbative model
calculations. This set of theoretical tools has been essential in
helping us to learn about QCD from the extensive experimental data,
having led in particular to numerous of measurements of the running
coupling, $\as$ \cite{Bethke:2002rv}, tests of the colour factors of
QCD \cite{SU3} and highly instructive studies concerning hadronisation
(for reviews, see \cite{Beneke,DasSalReview}).

Most of the investigations so far have been carried out for
observables that are sensitive to the deviation from two-jet
($1+1$-jet) structure in $\ee$ (DIS). This fairly straightforward
environment has been of considerable value, especially in developing
and refining our ideas about hadronisation. However the most stringent
tests of the understanding thus gained, as well as possible novel
extensions, are likely to be found in the context of multi-jet events,
including those at hadron-hadron colliders.

A difficulty in such studies is that the next-to-leading logarithmic
(NLL) resummed calculations, needed in the limit of only soft and
collinear emissions, have up to now usually been carried out by hand,
a tedious and error-prone procedure, which has to be repeated
separately for each new observable in each process. The complexity of
such calculations can increase substantially as one goes to multi-jet
processes \cite{eeKout}. 

This is to be contrasted with fixed-order calculations, usually
embodied in the form of a `fixed-order Monte Carlo' program (FOMC, \eg
\cite{NLOJET}), which given a subroutine for the observable, provides,
numerically, the leading and next-to-leading order (LO, NLO)
coefficients of the perturbative series. Though the internal
complexity of FOMCs increases for processes with extra legs, the
principles of their use remain identical regardless of the process.
This makes it possible, even for those not expert in higher order QCD
calculations, to obtain fixed-order predictions for arbitrary
observables, in a wide range of processes.

It is natural therefore to ask whether analogous programs could be
constructed for resummed calculations. To help explain a solution to
this question, proposed in \cite{caesar}, it is useful to define the
problem a little more precisely.  We will consider a general event
shape (or jet-resolution threshold), defined by some function
$V(q_1,\ldots,q_N)$ of the momenta $q_1,\ldots,q_N$ of the $N$
particles in an event. If the observable is such that it vanishes
smoothly in the limit of $n$ narrow jets, then we call it an
$(n+1)$-jet observable, and resummation will be needed in the $n$-jet
limit. The resummed calculations should provide results for the
probability $f(v)$ that the event shape has a value smaller than $v$.
For small $v$, this probability can often (exceptions are discussed
below) be written in the `exponentiated' form
\begin{equation}
  \label{eq:vProb-general}
  f(v) \simeq \exp\left[ L g_1(\as L) + g_2(\as L)
  + \as g_3(\as L) + \cdots\right]\,, \qquad\quad L = \ln \frac1v\,,
\end{equation}
where $Lg_1(\as L)$ contains LL terms, $\as^n L^{n+1}$, $g_2(\as L)$
the NLL ones and so forth. Usually NLL accuracy represents the state
of the art (though a related kind of resummation has recently been
performed to NNLL accuracy ($\as g_3$) for the $\ee$ energy-energy
correlation~\cite{deFlorian:2004mp}). 

In analogy with FOMCs, which calculate the LO and NLO contributions to
$f(v)$, the aim of an automated resummation program should be the
calculation (separately) of the resummed LL and NLL contributions, \ie
the $g_1(\as L)$ and $g_2(\as L)$ functions, free of any contamination
from potentially spurious higher orders, and in a form such that they
can be expanded, allowing matching to fixed-order
calculations.\footnote{This can be contrasted with the output from
  exclusive Monte Carlo event generators like Herwig \cite{Herwig} or
  Pythia \cite{Pythia}, which embody resummation for almost any
  observable via parton branching algorithms, and which provide an
  overall result for $f(v)$, generally correct to LL accuracy,
  sometimes to NLL, but for which it is difficult to guarantee the
  accuracy, extract separately the LL and NLL terms, perform matching
  to fixed order, and unambiguously separate the perturbative and
  non-perturbative effects.} %

The construction of an automated resummer, as least in our approach
\cite{caesar}, proceeds however in quite a different fashion from an
FOMC (usually just a Monte Carlo integrator with appropriate weights).
This is because, in order to separate out the structure of
(\ref{eq:vProb-general}) it is useful to have some analytical
understanding of the observable --- for example regarding its
behaviour in the presence of a soft and collinear emission, nearly
always of the form
\begin{equation}
  \label{eq:simple-second-time}
  V(\{{\tilde p}\}, k)=
  d_{\ell}\left(\frac{k_t}{Q}\right)^{a_\ell}
  e^{-b_\ell\eta}\, 
  g_\ell(\phi)\>,
\end{equation}
where $\{{\tilde p}\}$ are the hard momenta in the event and $k$ is a
soft emission that is collinear to the hard momentum (leg) with index
$\ell$, with a transverse momentum, rapidity and azimuthal angle
$k_t$, $\eta$ and $\phi$ relative to leg $\ell$.  For a large class of
observables, given the coefficients $a_\ell$ and $b_\ell$, the
function $g_1(\as L)$ has a known simple analytical form. So an
automated resummer can determine the LL terms by verifying the
parametrisation (\ref{eq:simple-second-time}) and determining the
coefficients.

This sounds remarkably simple, and indeed there is a `catch': one has
also to establish that a given observable belongs to the class for
which the LL terms truly are just determined by the values of the
$a_\ell, b_\ell$. Understanding precisely what that class is has been
one of the significant developments of \cite{caesar}, with the
introduction of a concept that we have called recursive infrared and
collinear (rIRC) safety. Normal infrared and collinear safety (IRC)
states that when an emission is very soft and/or collinear it should
have no effect on the observable. The recursive variant essentially
extends this condition to two-scale problems (as is relevant when
resumming) and states that given an ensemble of arbitrarily soft and
collinear emissions, the addition of further emissions of similar
softness or collinearity should not change the value of the observable
by more than a factor of order one.  Furthermore the addition of
relatively much softer or more collinear emissions (whether with
respect to the hard leg or one of the other emissions) should not
change the value of the observable by more than some power of the
relative extra softness or collinearity. While at first sight this
seems similar to normal IRC safety, a number of observables have been
found to be IRC safe but rIRC unsafe.  Recursive IRC safety is a
sufficient condition for the form (\ref{eq:vProb-general}) to hold.
While in principle one could also imagine resumming rIRC unsafe
observables, we are not aware of any cases where this has been done.

At NLL accuracy, it is not sufficient to know just the observable's
behaviour when there is a single soft and collinear emission ---
additionally one needs to know how the observable behaves when there
are multiple soft and collinear emissions of roughly similar hardness
(its `multiple-emission dependence'). Analytically this usually
requires an in-depth analysis of the observable. However, as was shown
in \cite{numsum}, there exists a general solution to the problem,
namely that at NLL all multiple-emission dependence can be taken into
account via a contribution $-\ln \cF(R')$ to $g_2$,
\begin{multline}
  \label{eq:cF-evshps-anyn}
  \cF(R') = \lim_{\epsilon\to0}
  \frac{\epsilon^{R'}}{R'}
    \sum_{m=0}^{\infty} \frac{1}{m!}
  \left( \prod_{i=1}^{m+1} \sum_{\ell_i=1}^n  R_{\ell_i}'
    \int_{\epsilon}^{1} \frac{d\zeta_i}{\zeta_i} 
    \int_0^{2\pi} \frac{d\phi_i}{2\pi}
  \right)   \delta\!\left(\ln \zeta_1\right)
  \times \\ \times
  \exp\left(-R'\ln \lim_{{\bar v}\to0} \frac{V(\{\tilde p\},\kappa_1(\zeta_1
      {\bar v}) , \ldots,
      \kappa_{m+1}(\zeta_{m+1}
      {\bar v}))}{\bar v}\right),
\end{multline}
where $R'(\as L)=-\partial_L L g_1(\as L)$ ($R'+1$ determines the
average number of `relevant' emissions), $R'_\ell$ is the part of $R'$
associated with leg $\ell$, and $\kappa_i(\bar v)$ is any soft
momentum collinear to leg $\ell_i$ that satisfies $V(\{\tilde
p\},\kappa_i(\bar v)) = \bar v$.  Since $\cF(R')$ can be calculated by
Monte Carlo integration, one eliminates the need for any analytical
study of the observable (the various limits can be taken numerically),
thus providing the main NLL element of an automated resummation.

One unsatisfactory issue that remains with the above approach is
connected with the problem of globalness. It was shown in \cite{NG1}
that for observables that are sensitive to emissions only in a
restricted angular region (for example a single jet), additional
classes of NLL terms arise, termed non-global logarithms (NGLs), so
far calculated only in the large-$\NC$ limit (though see also
\cite{weigert}). Such NGLs depend non-trivially on the details of the
boundaries of the restricted region in which the observable is
sensitive to emissions, and these boundaries can be quite complex.
Accordingly, for the time being, we have chosen to consider only the
simpler case of (continuously) global observables, which have the
characteristic that the $d_\ell$ are all non-zero and the $a_\ell$ are
all equal. These observables' resummations are free of NGLs.

The resulting resummation program is called \caesar, the Computer
Automated Expert Semi-Analytical Resummer, and many results obtained
with it can be found at
\href{http://qcd-caesar.org}{\texttt{http://qcd-caesar.org}}, for
$\ee$, DIS and hadron-hadron collisions.

Particularly topical are hadron-collider dijet event-shapes, the
subject also of recent experimental \cite{D0Thrust} and theoretical
fixed-order \cite{Nagy03} work. These are of interest for a variety of
reasons, ranging from the possibility of `standard' measurements such
as $\as$, to more novel studies examining for example the underlying
event with techniques such as those discussed in \cite{KoutZ0}, or the
soft (colour evolution) anomalous dimension matrices of
\cite{Sterman4Legs}. The event shape studied in \cite{D0Thrust,Nagy03}
(a transverse thrust) has the drawback that its resummation involves
NGLs, and so is beyond the current scope of \caesar. To allow more
straightforward compatibility with resummed calculations, it is
instead preferable to specifically design a set of global dijet event
shapes (and jet rates). This has been the subject of work in
\cite{BSZhh}. One of the main practical issues discussed there is the
question of how to reconcile the requirement of globalness with the
limited experimental detector coverage at large rapidities.
Fortunately this problem can be circumvented in a variety of ways.
This opens the road at hadron-colliders to full resummed studies
matched with fixed order calculations \cite{NLOJET}, with the prospect
of significantly extending the very fruitful studies carried out in
$\ee$ and DIS over the past years.

\textbf{Acknowledgements.} The work discussed here has been carried out
in collaboration with Andrea Banfi and Giulia Zanderighi. I with to
thank them also for helpful comments on this write-up.



\begin{thebibliography}{99}

\bibitem{Bethke:2002rv} S.~Bethke,
Nucl.\ Phys.\ Proc.\ Suppl.\  {\bf 121} (2003) 74
[hep-ex/0211012].

\bibitem{SU3}
S.~Kluth et al.,
Eur.\ Phys.\ J.\ C {\bf 21} (2001) 199  and references therein.

\bibitem{Beneke}
M.~Beneke,
Phys.\ Rept.\  {\bf 317} (1999) 1.

\bibitem{DasSalReview}
M.~Dasgupta and G.~P.~Salam,
J.\ Phys.\ G {\bf 30} (2004) R143
[hep-ph/0312283].


\bibitem{eeKout} 
A.~Banfi, G.~Marchesini, Yu.~L.~Dokshitzer and G.~Zanderighi,
JHEP {\bf 0007} (2000) 002 
[hep-ph/0004027]; 
JHEP {\bf 0105} (2001) 040
[hep-ph/0104162].

\bibitem{NLOJET}
Z.~Nagy,
Phys.\ Rev.\ Lett.\  {\bf 88} (2002) 122003
[hep-ph/0110315].


\bibitem{caesar}
A.~Banfi, G.~P.~Salam and G.~Zanderighi,
Phys.\ Lett.\ B {\bf 584} (2004) 298
[hep-ph/0304148]; hep-ph/0407286.

\bibitem{deFlorian:2004mp}
D.~de Florian and M.~Grazzini,
hep-ph/0407241.

\bibitem{Herwig}
G.~Abbiendi  {\it et al.},
Comput.\ Phys.\ Commun.\  {\bf 67} (1992) 465.
G.~Corcella {\it et al.},
JHEP {\bf 0101} (2001) 010
[hep-ph/0011363].

\bibitem{Pythia}
T.~Sj\"ostrand,
Comput.\ Phys.\ Commun.\  {\bf 82} (1994) 74;
T.~Sj\"ostrand {\it et al.},
Comput.\ Phys.\ Commun.\  {\bf 135} (2001) 238;
[hep-ph/0010017].

\bibitem{numsum} A.~Banfi, G.~P.~Salam and G.~Zanderighi,
JHEP {\bf 0201} (2002) 018
[hep-ph/0112156].

\bibitem{NG1} 
M.~Dasgupta and G.~P.~Salam,
Phys.\ Lett.\ B {\bf 512} (2001) 323
[hep-ph/0104277];
JHEP {\bf 0203} (2002) 017
[hep-ph/0203009].

\bibitem{weigert}
H.~Weigert,
Nucl.\ Phys.\ B {\bf 685} (2004) 321
[hep-ph/0312050].

\bibitem{D0Thrust}
I.~A.~Bertram  [D0 Collaboration],
Acta Phys.\ Polon.\ B {\bf 33} (2002) 3141.

\bibitem{Nagy03}
Z.~Nagy,
Phys.\ Rev.\ D {\bf 68} (2003) 094002
[hep-ph/0307268].

\bibitem{KoutZ0}
A.~Banfi, G.~Marchesini, G.~Smye and G.~Zanderighi,
JHEP {\bf 0108} (2001) 047
[hep-ph/0106278].


\bibitem{Sterman4Legs}
J.~Botts and G.~Sterman,
Nucl.\ Phys.\ B {\bf 325} (1989) 62;
N.~Kidonakis and G.~Sterman,
Nucl.\ Phys.\ B {\bf 505} (1997) 321
[hep-ph/9705234];
N.~Kidonakis, G.~Oderda and G.~Sterman,
Nucl.\ Phys.\ B {\bf 531} (1998) 365
[hep-ph/9803241];
G.~Oderda,
Phys.\ Rev.\ D {\bf 61} (2000) 014004
[hep-ph/9903240].

\bibitem{BSZhh} A.~Banfi, G.~P.~Salam and G.~Zanderighi, 
hep-ph/0407287. 
\end{thebibliography}
\end{document}